\documentclass[12pt]{article}
\usepackage{graphicx}
\begin{document}
version 14.3.2014
\begin{center}
{\Large What is the Real Structure of Fundamental Forces? }   
\vspace{0.3cm}

H.P. Morsch \\ HOFF, Brockm\"ullerstr.~11,
D-52428 J\"ulich, Germany\\  E-mail:
h.p.morsch@gmx.de 
\end{center}

\begin{abstract}
Based on general considerations, the Standard Model of particle physics 
with its extensions (SM) can be ruled out as a valid theory of 
fundamental forces: it requires far too many parameters, which are not 
determined from first principles. The only way to uncover   
the real structure of elementary forces is to resort to a complete and 
fundamental theory {\bf without} external parameters. 

This requirement is fulfilled in a finite theory, based on an
extension of the QED Lagrangian by boson-boson coupling, with massless
elementary fields, 
charged and neutral elementary fermions (quantons) and gauge bosons. 
In this description the observed bound states of nature, hadrons, leptons, 
atoms and gravitational objects are understood as stationary systems of
quantons bound by electric or magnetic forces. Other fields do not exist.

PACS/ keywords: 11.15.-q, 12.10.-g/ Description of elementary forces
in a complete (fundamental) theory, based on a Lagrangian with Maxwell term,
boson-boson coupling and massless elementary fermions (quantons). Bound states
of nature described by systems bound by electric or magnetic forces. 
\end{abstract}

The study of the fundamental forces of nature allows to gain
insight into the early development of the universe. But this requires a close
theoretical description, in which all model assumptions can be varified and the
needed parameters are derived from first principles. 
Up to now the Standard Model of particle physics~\cite{PDG} (SM), established
about 40 years ago, represents the state of our knowledge. It is a heuristic
model constructed from relativistic first order quantum field theories,
different for each fundamental force (gravitation is not included in the SM,
since a quantum theory of gravity is not well established), with a number of
fields of different symmetry. 
Further, the understanding of the masses of elementary
fermions requires background (Higgs) fields, but for neutrino masses in
addition the postulation of heavy Majorana neutrinos is needed.  The flavour 
degree of freedom observed in hadrons and leptons is
understood in supersymmetric extensions of the SM. In particular, the minimal
supersymmetric SM (MSSM) has been favoured, since it allows to understand the
hierarchy problem between the weak and gravitational forces. Further, in this
model the running couplings of the electric, weak and strong forces meet on the
grand unification scale $\Lambda_{GUT}=10^{16}$ GeV. However, in the MSSM
supersymmetric particles have been 
predicted in the few hundred GeV region, which have not been found
experimentally. These negative results cast serious doubts on the 
validy of the SM including its proposed extensions. Since it is not very
probable that other grand unified theories can resolve these problems, it is
time to examine what model assumptions may be unrealistic or even wrong. 

A serious point of critics of the SM is that very different theories are
assumed for each fundamental interaction. This gives rise to a large  
complexity\footnote{It should have been questioned, whether nature is not more
  economic and efficient to organise the creation of hadrons and leptons --
  the constituents of matter -- in a much simpler way.} in the weak
and strong interaction sectors with fields of different symmetry, 12
elementary fermions and quite a few massless but also massive gauge bosons. 
The consequence of this construction is that a large number of
parameters have to be adjusted (the masses of elementary particles, mixing
parameters, couplings between fields, etc.). An understanding of most of these 
parameters (those, which cannot be determined from first principles) requires 
further theoretical explanations\footnote{SM parameters should not just be
  adjusted to experimental data, they have to be understood!} with
still other parameters, which again have to be understood. This leads to an
endless chain of related parameters without clearly defined solution, proving
that the assumptions in the SM are erroneous or unrealistic. 

Examples of such parameters are mass and flavour of elementary particles.
Their explanation requires additional background (Higgs) and
supersymmetric fields. The couplings of the Higgs-fields to massless fermions
have to be considered as additional parameters, which could be understood only
by a further theory with additional rather uncertain parameters. 
Even worse, supersymmetry gives rise to many combinations of
fermion and boson fields with mixing parameters of large
uncertainties, which cannot be resolved from first principles. This indicates
clearly that these explications cannot be correct, but also that a realistic
theory should have a much simpler structure with very few parameters. 

The only way to avoid the above problems is to resort to a
complete and fundamental theory, in which all parameters are determined from
first principles. Such a theory should have only a minimum of fields and thus
significantly less degrees of freedom than the SM. Also in such a theory
relativistic bound states have to be described explicitely. 

If we inspect the structure of the SM, even quantum electrodynamics (QED), the
best established part of the SM, cannot be regarded as a closed and
fundamental theory, since the coupling constant $\alpha_{QED}\sim$1/137 is an
external parameter determined from experiment. However, the precise prediction
of spin properties suggests that QED is close to a fundamental theory. Only 
the divergencies for $r\to 0$ and $\infty$ appear to be in conflict with
nature, which is known to develop in a smooth and finite way. Because of this,
a complete and fundamental theory is expected to be similar to QED, but with
a structure more complex than a first order gauge theory. 

During the last years a theoretical framework has been
developed~\cite{Mo1,Mo2,Mo3}, based on a second order extension of QED, which
shows all features expected of a fundamental theory: all needed parameters can
be determined by self-consistency conditions, thus showing completeness. With
massless elementary 
bosons and fermions (quantons) this model shows a coupling to the vacuum with
average energy $E_{vac}=0$. As a crucial test, by applying this model to the 
binding of light atoms~\cite{Mo3}, the self-consistently deduced coupling
constant $\alpha$ was found to be consistent with $\alpha_{QED}$. This is
needed, since the Coulomb potential\footnote{The Coulomb potential has to be
  considered as effective potential, since the restoring force is taken over
  by the potential itself.} yields eigenvalues in agreement with the
experimental spectrum. 

It has to be noted that in the past many different Lagrangians have been
studied, with the general conclusion that higher order theories should be
discarded, because they could lead to unphysical
solutions~\cite{highd}. However, the important difference to the higher order
theories discussed in ref.~\cite{highd} is that in the present formalism the
Lagrangian is gauge invariant and non-physical solutions can be eliminated by
strict geometric, mass-radius and energy-momentum relations. Further, only in
a second order formalism relativistic bound states can be generated, which are
stable, see the discussion below. 

The Lagrangian has been used in the form 
\begin{equation}
\label{eq:Lagra}
{\cal L}=\frac{1}{\tilde m^{2}} \bar \Psi\ i\gamma_{\mu}D^{\mu}
D_{\nu}D^{\nu}\Psi\ -\ \frac{1}{4} F_{\mu\nu}F^{\mu\nu}~,   
\end{equation}
where $\tilde m$ is a mass parameter and $\Psi$ a two-component massless
fermion field (with charge and neutral part), $\Psi=(\Psi^+, \Psi^o)$ and
$\bar \Psi= (\Psi^-, \bar \Psi^o)$.  
Vector boson fields $A_\mu$ with coupling $g$ to fermions are contained in the
covariant derivatives $D_{\mu}=\partial_{\mu}-i{g} A_{\mu}$. 
The second term of the Lagrangian represents the Maxwell term with
Abelian field strength tensors $F^{\mu\nu}$ given by $F^{\mu\nu}= 
\partial^{\mu}A^{\nu}-\partial^{\nu}A^{\mu}$, which gives rise to both electric 
and magnetic effects. 

By inserting $D^{\mu}=\partial^{\mu}-i{g} A^{\mu}$ and
$D_{\nu}D^{\nu}=\partial_{\nu}\partial^{\nu}
-ig(A_{\nu}\partial^\nu+\partial_\nu A^\nu) -g^2 A_\nu A^\nu$ in
eq.~(\ref{eq:Lagra}), the first term of ${\cal L}$ gives rise to a number of
different terms, which contain boson and fermion fields and/or their
derivatives, see ref.~\cite{Mo2}. All terms, which contain the
derivative of the fermion field $\partial^{\nu} \Psi$, are related to a rather
complex dynamics of the system. For stationary solutions only two terms of the
Lagrangian contribute
\begin{equation}
\label{eq:L2}
{\cal L}_{2g} =\frac{-ig^{2}}{\tilde m^{2}} \ \bar \Psi \gamma_{\mu}
[A^\mu \partial_{\nu} A^\nu]~\Psi \ 
\end{equation}
and 
\begin{equation}
\label{eq:L3}
{\cal L}_{3g} =\frac{ -g^{3}\ }{\tilde m^{2}} \ \bar \Psi \gamma_{\mu}
[A^\mu A_\nu A^\nu]~\Psi \ .
\end{equation}
The gauge condition $\partial_\mu A^\mu=0$ used for first order Lagrangians
is replaced in the present case by $\partial(\partial_\nu A^\nu)=0$. 

From the Lagrangians~(\ref{eq:L2}) and (\ref{eq:L3}) fermion matrix elements 
have been evaluated. This has been found to be a reliable
method to generate bound state potentials, see refs.~\cite{Mo2,Mo3}. 
Using $\alpha=g^2/4\pi$ and fermion wave functions $\psi(p)=\frac{1}{\tilde
  m^{3/2}} \Psi(p_i)\Psi(k)$, matrix elements are obtained of the form 
\begin{equation}
\label{eq:M2}
{\cal M}_{2g} =\frac{\alpha^{2}}{\tilde m^8} \bar \psi(p')\gamma_\mu
A^{\mu}(q)~(\partial_\nu A^{\nu}(q))(\partial_\sigma
A^{\sigma}(q))~\gamma_\rho A^{\rho}(q) \psi(p)\ 
\end{equation}
and
\begin{equation}
\label{eq:M3}
{\cal M}_{3g} = \frac{-{\alpha}^{3}}{\tilde m^8}~\bar
\psi(p')\gamma_\mu A^\mu(q)~A_\nu(q)A^\nu(q)~
A_\sigma(q)A^\sigma(q)~\gamma_\rho A^\rho(q) \psi(p) \ , 
\end{equation} 
Theses matrix elements have a structure more complex than obtained in a first
order theory~\cite{matrix}. But this is necessary to describe relativistic 
bound states in a fundamental approach: only by interactions between fermions 
a stable system cannot be generated; the additional boson fields on the right
and left of ${\cal M}_{2g}$ and ${\cal M}_{3g}$ have to provide the restoring
force needed to stabilise the system. 

From these matrix elements bound state potentials can be deduced. Following
the derivations in refs.~\cite{Mo2,Mo3} ${\cal M}_{2g}$ and ${\cal M}_{3g}$ 
can be simplified, using (analogue to the fermion wave functions) normalised
boson (quasi) wave functions $W_{\mu}^\nu(q)= \frac{1}{\tilde m} A_\mu(q)
A^\nu(q)$ and a boson-exchange interaction $V_{\mu}^\nu(q)= W_{\mu}^\nu(q)$
($\mu\neq\nu$). 
Further, by an equal time requirement the fermion and boson vectors can be 
reduced by one dimension, yielding boson wave functions\footnote{with
  dimension $[GeV]$.} of scalar and vector structure $w_{s}(q)$ and $w_{v}(q)$
and an interaction potential $v_{v}(q)= w_{v}(q)$. Going to r-space the 
fermion matrix elements~(\ref{eq:M2}) and (\ref{eq:M3}) can be written by 
\begin{equation}
\label{eq:P2g}
{\cal M}^f_{2g} = \bar \psi(r)\ V_{2g}(r)\ \psi(r)\ 
\end{equation}
and
\begin{equation}
\label{eq:M3f}
{\cal M}^f_{3g} = \bar \psi(r)\ V_{3g}^{s,v}(r)\ \psi(r) \ ,
\end{equation}
where the bosonic potentials $V_{2g}(r)$ and $V_{3g}^{s,v}(r)$ are given by
\begin{equation}
V_{2g}(r)= \frac{\alpha^2 (\hbar c)^2 F_{2g}}{4\tilde m}\ \Big
(\frac{d^2 w_s(r)}{dr^2} + 
  \frac{2}{r}\frac{d w_s(r)}{dr}\Big )\frac{1}{\ w_s(r)}+E_o\  
\label{eq:vb}
\end{equation}
and 
\begin{equation} 
\label{eq:vqq}
V^{s,v}_{3g}(r)= -\frac{\alpha^3 \hbar c}{\tilde m} \int dr'\ 
w_{s,v}(r')\ v_v(r-r')\ w_{s,v}(r')~.   
\end{equation}
The factor $F_{2g}$ in eq.~(\ref{eq:vb}) is due to the Fourier 
transformation of the boson kinetic energy and $E_o$ the energy 
of the lowest eigenstate. A connection to the vacuum is made by 
assuming $E_o=E_{vac} = 0$. $V_{2g}(r)$ shows a quite 
linear rise towards larger radii, leading to confinement of the 
system; therefore it can be identified with the
confinement potential required in hadron potential models~\cite{qq}.
The potential $V^{s,v}_{3g}(r)$ can be considered as boson matrix element, 
in which the wave functions $w_{s,v}(r)$ are bound in the potential $v_v(r)$.

From the structure of the fermion matrix element in eqs.~(\ref{eq:M3f}) and
(\ref{eq:vqq}) one can see that there are two $q\bar q$ states (with quantum
numbers $J^\pi=1^-$) with scalar and vector boson wave functions
$w_{s,v}(r)$ and corresponding fermion wave functions $\psi_{s,v}(r) \sim
w_{s,v}(r)$. Further, there are two $q\bar q$ p-states (with quantum numbers
$J^\pi=0^+$) with similar wave functions, see ref.~\cite{Mo3,Mo4}.

The fermion wave functions have to be orthogonal, leading to the constraint
\begin{equation}
\label{eq:ortho}
\int r^2dr~\psi_s(r) \psi_v(r)=\int r^2dr~w_s(r) w_v(r)=<r_{w_s,w_v}>=0 \ .
\end{equation}
To satisfy this condition, $w_{v}(r)$ can be written in the form of a
p-wave function 
\begin{equation}
\label{eq:spur}
w_{v}(r) = w_{v,o}~[w_s(r)+\beta R\ \frac{d w_s(r)}{dr}]~,
\end{equation}
where $w_{v,o}$ is obtained from the normalisation $2\pi \int r dr\ w_v^2(r)
=1$ and $\beta R$ determined by $\beta R=-\int r^2dr~w_s(r)/\int
r^2dr~[dw_s(r)/dr]$. 
Interestingly, orthogonality gives rise to another quite natural condition for
the deepest bound state, requiring that the interaction takes place inside the
bound state volume of $w_s^2(r)$. This leads to 
\begin{equation}
\label{eq:conr}
|V^v_{3g}(r)| \sim {c}\ w^2_s(r)  \ . 
\end{equation}

The conditions~(\ref{eq:ortho})-(\ref{eq:conr}) lead to a
boson wave function $w_s(r)$ of the form 
\begin{equation}
\label{eq:wf}
w_s(r)=w_{s,o}\ exp\{-(r/b)^{\kappa}\} \ , 
\end{equation} 
where $w_{s,o}$ is fixed by the normalisation $2\pi \int
r dr\ w_s^2(r) =1$. The parameters $b$ and $\kappa$ have to be determined by
boundary conditions as discussed below. Different flavour states are
obtained by solutions with different slope parameter $b$, which are
constrained by a vacuum potential sum rule, see e.g.~ref.~\cite{Mo3}.
The interaction $v_v(r)$ is given by $v_v(r)=\hbar c~w_v(r)$.

Binding energies have been calculated within
the Hamiltonian formalism by using
the virial theorem in the form $4\pi[\int r^2dr~\psi^2(r)V_{ng}(r)
  -\frac{1}{2}\int r^3dr~\psi^2(r)\frac{d}{dr}V_{ng}(r)]=E^{ng}_f$, where
$\psi(r)$ are fermion wave functions with a form similar to the boson wave
functions in eqs.~(\ref{eq:spur}) and (\ref{eq:wf}). In addition,
$V^{s}_{3g}(r)$ can be interpreted as bound state of bosons. Its binding
energy $E_g$ has been calculated by the corresponding form 
$2\pi[\int rdr~w_s^2(r)v_v(r) -\frac{1}{2}\int
  r^2dr~w_s^2(r)\frac{d}{dr}v_v(r)] =E_g$.
For massless fermions the mass of the system $M^{s,v}$ is given by the
absolute binding energies in $V_{2g}(r)$ and $V^{s,v}_{3g}(r)$, yielding
$M^{s,v}=-E^{3g}_{f_{s,v}}+E^{2g}_f$, while the reduced mass is given by
$\tilde m=M^{s}/2$.  

In order to make a detailed evaluation of the potentials other constraints are
needed, which connect the coupling constant $\alpha$ to the shape parameters
$\kappa$ and $b$ and the mass of the system. 
If the mass is adjusted to that known experimentally, only two conditions
are needed for a complete determination of all parameters. This has been done
in most cases. However, by applying in addition a vacuum potential sum rule,
also the masses of different solutions are
constrained, thus leading to a fully complete description. 
 
The first condition is energy-momentum conservation, important for  
relativistic bound states.  For $(q\bar q)$ as well as $(p~e^-)$ and $(e^+
e^-)$ systems bound by electric interactions, the
negative fermion and boson binding energies $E_f^s$ and $E_g$ have to be
compensated by their root mean square momenta $<q^2_{v}>=\int dq~q^3~v(q)/\int
dq~q~v(q)$ in the potentials $V_{3g}(r)$ and $v_{v}(r)$, respectively
\begin{equation}
<q^2_{V_{3g}}>+<q^2_{v_v}> = (E_f^s+E_g)^2 \ . 
\label{eq:massq}
\end{equation}
For more complex $(q\bar q)^n$ and $(q\bar q)^nq$ systems somewhat different
energy-momentum relations are needed. Of particular interest are $(q\bar
q)^nq$ states with opposite fermion momenta $\sum_i <q^2_{\psi_i}>=0$. In this
case the motion of fermions give rise to magnetic interactions and thus to
magnetic binding. This leads to leptonic bound states~\cite{Mo5} with an
energy-momentum relation of the form 
\begin{equation}
<q^2_{v_{v}}> ({v}/{c})^2 = (E_f^s+E_g)^2 \ , 
\label{eq:mqmag}
\end{equation}
where $({v}/{c})^2$ is the relative motion of fermions, $({v}/{c})^2
<6~10^{-19}$ for leptons, see ref.~\cite{Mo5}.  

A mass-radius condition can be derived from the confinement
potential~(\ref{eq:vb}), see ref.~\cite{Mo2}, which leads to 
\begin{equation}
Rat_{conf}=\frac{(\hbar c)^2}{\tilde m^2 <r^2_{w_s}>} =1 \ .
\label{eq:ravb}
\end{equation}
For magnetic binding eq.~(\ref{eq:ravb}) is multiplied by a factor of
  $({v}/{c})^2$. In this case p-wave states are not stable, leading only to
one lepton bound state (for each flavour system). 

The last ambiguity between $\alpha$, $\kappa$, $b$ and the mass of the bound
state can be removed by a vacuum potential sum rule. Here a
vacuum potential $v_{vac}(q)$ is assumed, which should be equal to the sum
of all individual interactions $v_v^i(q)$. This is expected to have a simple
form $\sim 1/q^n$ with a cut-off function $f_{cut}(q)$ to make $v_{vac}(q)$
finite. This leads to  
\begin{equation}
v_{vac}(q)= f_{cut}(q)~1/q^n =\sum_i v_v^i(q) \ 
\label{eq:vvac}
\end{equation}
and allows to determine $v_{vac}(q)$ by the sum of all individual bound
state solutions. In turn, $v_{vac}(q)$ can be used to check the mass assumptions
of the individual bound states. Further, hypothetical ground state wave
functions can be defined, which allows to check the consistency of this
approach. For atomic states this procedure yields very consistent results, see
ref.~\cite{Mo3}.

The general structure of the bound state solutions is shown in fig.~1. In the
upper part the radial dependence of the interaction $v_v(r)$ is compared to
the $1/r$ dependence of the Coulomb potential. In the middle part the radial
dependence of boson density $w^2_s(r)$ and potentials $V_{3g}^{s,v}(r)$ is
shown, which indicates that relation (\ref{eq:conr}) is well fulfilled. In the
lower part the confinement potential $V_{2g}(r)$ is given, which has the
typical linearly increasing behaviour for larger radii, as needed in empirical
hadron structure models~\cite{qq}. The same behaviour of boson densities and
potentials has been found for all systems studied, therefore the
horizontal and vertical scales are given only in relative units. 

Of large importance, by using boundary conditions for electric and magnetic
binding, systems of very different radii are obtained. Whereas for hadrons
(bound electrically) root mean square radii between $4~10^{-3}$ and 0.5 fm
have been extracted~\cite{Mo2,Mo4}, for leptons (bound magnetically) root mean
square radii smaller than $10^{-8}$ fm have been found~\cite{Mo5}. This
indicates that the decay of hadrons to leptons is very weak, as found
experimentally.

The decay width of the systems in question is related to the
dynamical structure of the potentials, given by kinetic energy distributions 
$T_{2g}(q)$ and $T_{3g}(q)$.  $T_{2g}(q)$ is given by the Fourier transformed
potential $V_{2g}(r)$, whereas $T_{3g}(q)$ is given by the Fourier transformed 
kinetic energy $T_{3g}(r)$ given by
\begin{equation}
T_{3g}(r)=\frac{1}{2} <r^2> 
(d^2 V_{3g}(r)/dr^2 + \frac{2}{r}d V_{3g}(r)/dr) \ .
\end{equation}
Results on $T_{2g}(q)$ and $T_{3g}(q)$ are given in fig.~2. Although in many
cases the confinement potential $V_{2g}(r)$ is much weaker than $V^{s,v}_{3g}(r)$, 
this potential gives rise to a very pronounced energy distribution. 
From these distributions all decay properties and branching ratios can be
deduced by calculating the overlap of $T_{2g}(r)$ and $T_{3g}(r)$ for all
systems involved. 
Intestingly, for mesonic $q\bar q$ systems the confinement potential gives rise
to a sharp peak consistent with the experimental width, see the discussion in
ref.~\cite{Mo4}, whereas the boson-exchange potentials $V_{3g}(r)$ yield
very wide structures, in the case of $V^s_{3g}(r)$ a peak of 
$\sim$2 GeV width for $J/\psi$ and $\sim$70 GeV width for the top system.    

------

Applications of the present formalism have shown that hadrons are
well described as states bound by electric forces, see ref.~\cite{Mo2,Mo4},
whereas leptons represent systems bound by magnetic forces, see
ref.~\cite{Mo5}. This underlines the inherent symmetry of electric and magnetic
forces in Maxwell's theory. These results confirm the general conclusion 
drawn above that in the SM both the strong and weak forces are incorrectly  
assumed with too many parameters. Concerning the strong interaction,
confinement is a basic property of all relativistic bound states, see fig.~1;
therefore 'colour confinement' in a theory of the strong interaction is
redundant, making a non-Abelian colour theory as quantum chromodynamics (QCD)
dispensable.  
With respect to the weak interaction, magnetic forces have been misinterpreted
in the SM as heavy-boson exchange. Both, magnetic and heavy-boson exchange
forces give rise to similar properties, extremely small radii and strength,
but the latter requires further Higgs fields to make the theory gauge
invariant.  
Finally, there is no basis for supersymmetry: the flavour degree of freedom is
characterised in the present formalism by systems of different slope parameter
$b$.

The implications are clear: apart from elementary fermion and boson fields 
all other fields assumed in the SM are unrealistic. This excludes
colour, heavy boson, Higgs, supersymmetric and all other exotic fields. 
Consequently, the experimental states $W^\pm$(80.4 GeV) and $Z$(91.2 GeV) have
to be reinterpreted as $(q\bar q)^nq$ and $(q\bar q)^n$ states, respectively. 
Actually, $Z$(91.2 GeV) has to be regarded~\cite{Mo4} as low mass top $q\bar
q$-state, with a width in 
agreement with experiment. Further, the new state\footnote{misinterpreted as
  SM Higgs-boson.} with a mass of 126 GeV, discovered~\cite{resnew} recently 
at LHC, has to be identified with a scalar $q\bar q$-state (of p-wave
structure), found with exactly this mass, see ref.~\cite{Mo4}. A second scalar
$q\bar q$-state is predicted 
with a mass of 41 GeV, which should be found in high energy experiments. 
 
A further strict consequence of the present formalism is that gravitation
has to be understood also by electromagnetic forces. 
Strong support for this conclusion
comes from the analysis of neutrinos, see ref.~\cite{Mo5}, indicating that  
a hierarchy problem between 'weak' and gravitational forces does not exist, if
we assume that gravitation is due to magnetic interactions of particles in
matter. The present formalism excludes definitely dark matter in form of exotic
particles (the rotation of galactic systems, considered as the best evidence
for dark matter, has been well described in the present approach, see
ref.~\cite{Mo1}). Finally, the present formalism has to be
considered as the proper theory of quantum gravity. For more detailed studies
other solutions of the Lagrangian~(\ref{eq:Lagra}) may be needed, which go
beyond the evaluation of matrix elements.  
\vspace{0.6cm}

In conclusion, by using general arguments the SM with its complex structure 
cannot be considered as a valid theory of fundamental forces. The 
only method to find a correct solution is to develop a 
complete and fundamental theory, which has no external parameters.  
Such a theory has been constructed, based on a second order extension of QED,
with the result that {\bf all elementary forces} have to be described by
electromagnetic interactions only. Apart from basic fermion and boson fields,
the existence of all other SM fields can be ruled out.
In particular, supersymmetric particles and dark matter in form of exotic 
particles do not exist.   

For cosmology a new scenario arises: we have to conclude that the
universe emerged out of the absolute vacuum of fluctuating boson fields with
average energy $E_{vac}=0$. During overlap of two boson fields $q\bar
q$-pairs could be created, which have been confined to form hadrons and
leptons (for this process $V_{2g}(r)$ is of prime importance). By
electric forces neutral matter was created, which bound and compressed 
by gravitation to a dense cosmic bound state. 
Above a critical density the largest part
of matter annihilated (the detailed mechanism have still to be worked out) and 
gave a big radial impact (Big Bang) to the remaining matter, mainly in form of
bound ($p~e^-$)-pairs, resulting in an
expansion of the system. By perpetual decrease of the binding energy of the
cosmic bound state the expanding matter gained and still gains extra momentum,
leading in this way to an (ever) increasing expansion of the universe, without 
dark matter in form of exotic particles.  

After one century of modern physics with the development of very complex
theories for fundamental forces, it may be
surprising that all basic bound states of nature with a strikingly different
phenomenology can be understood as stationary states bound by electric or
magnetic forces. However, this is in perfect agreement with the general rules
of nature, effectiveness and formal logic, by which complex systems should 
evolve from the simplest structure (the vacuum) by the simplest
possible mechanisms. \\  

For fruitful discussions, direct help in the derivation of the formalism and
general support the author is indepted to many colleagues, with a list of
their names given in a final publication.

\begin{figure}
\centering
\includegraphics [height=18cm,angle=0] {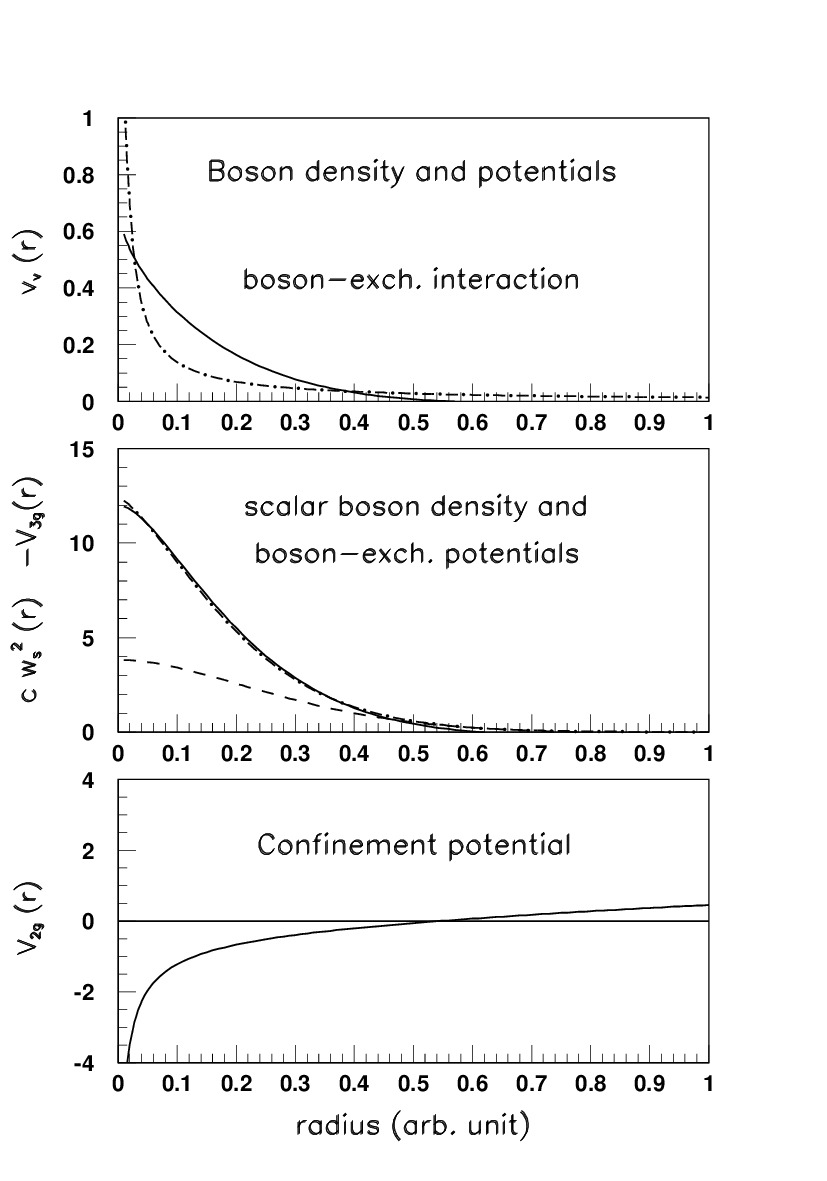}
\caption{General structure of boson density and potentials.
  \underline{Upper part:} Relative interaction $v_v(r)$ in comparison with the
  Coulomb potential $(\sim 1/r)$ given by dot-dashed line.
 \underline{Middle part:} Boson density $w_s^2(r)$ (dot-dashed line) and
 boson-exchange potentials $|V^{s,v}_{3g}(r)|$ given by dashed and solid
 lines, respectively. 
 \underline{Lower part:} Deduced confinement potential $V_{2g}(r)$. }    
\label{fig:g1exep}
\end{figure} 

\begin{figure}
\centering
\includegraphics [height=18cm,angle=0] {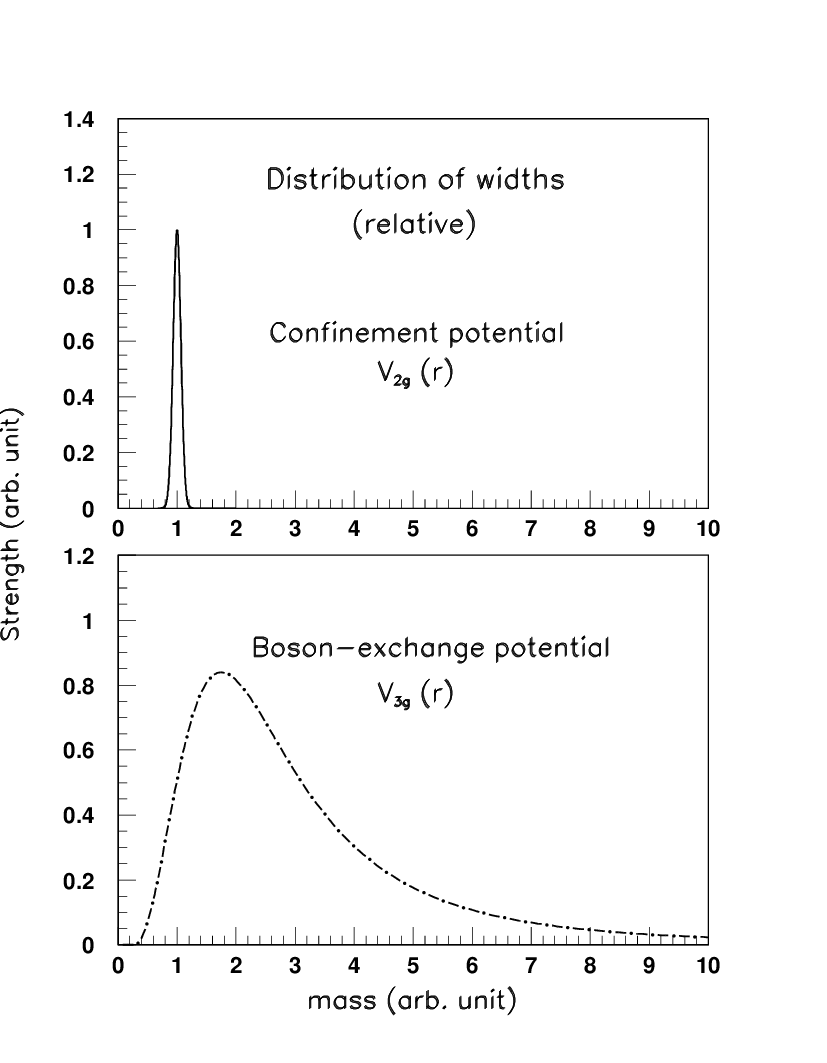}
\caption{Relative distributions of widths deduced for the confinement
  potential (upper part) and the boson-exchange potential $V^s_{3g}(r)$ (lower
  part). }    
\label{fig:g1exelec}
\end{figure} 

\end{document}